\documentclass[onecolumn,
               preprintnumbers,
               superscriptaddress,
               numbers,
               floatfix,
               nofootinbib,
               showpacs,
               colorlinks,
               linkcolor=blue,   
               citecolor=blue]{revtex4-2}

\usepackage{graphicx}
\graphicspath{{Figures/}}
\usepackage{physics}   
\usepackage{adjustbox} 
\usepackage{wrapfig}
\usepackage{float}
\usepackage[utf8x]{inputenc}
\usepackage{amsmath}
\usepackage{amssymb}
\usepackage{orcidlink}
\usepackage{amsthm}
\usepackage{mathtools}
\usepackage{color}
\usepackage{multirow}
\usepackage{tabularx}
\usepackage{comment}
\usepackage{soul}
\usepackage{siunitx}
\usepackage{booktabs}
\DeclareSIUnit\bar{bar}

\begin{document}

\title{WISPFI Experiment: Prototype Development}
\affiliation{Institut für Experimentalphysik, Universität Hamburg, Luruper Chaussee 149, D-22761 Hamburg, Germany}

\author{Josep Maria Batllori\orcidlink{0000-0003-0555-8970}}
\affiliation{Institut für Experimentalphysik, Universität Hamburg, Luruper Chaussee 149, D-22761 Hamburg, Germany}

\author{Michael Frosz\orcidlink{0000-0002-8857-0029}}
\affiliation{Max Planck Institute for the Science of Light, Erlangen, 91058, Germany}

\author{Dieter Horns\orcidlink{0000-0003-1945-0119}}
\affiliation{Institut für Experimentalphysik, Universität Hamburg, Luruper Chaussee 149, D-22761 Hamburg, Germany}

\author{Marios Maroudas\orcidlink{0000-0003-1294-1433}}
\affiliation{Institut für Experimentalphysik, Universität Hamburg, Luruper Chaussee 149, D-22761 Hamburg, Germany}

\begin{abstract}

Axions and axion-like particles (ALPs) are well-motivated dark matter (DM) candidates that couple to photons in external magnetic fields. The parameter space around $m_a \sim \SI{50}{\milli\eV}$ remains largely unexplored by haloscope experiments. We present the first prototype of WISP Searches on a Fiber Interferometer (WISPFI), a table-top, model-independent scheme based on resonant photon--axion conversion in a hollow-core photonic crystal fiber (HC-PCF) integrated into a Mach--Zehnder interferometer (MZI). Operating near a dark fringe with active phase-locking, combined with amplitude modulation, the interferometer converts axion-induced photon disappearance into a measurable signal. A \SI{2}{\watt}, \SI{1550}{\nano\meter} laser is coupled into a \SI{1}{\meter}-long HC-PCF placed inside a $\sim$\SI{2}{\tesla} permanent magnet array, probing a fixed axion mass of $m_a \simeq \SI{49}{\milli\eV}$ with a projected sensitivity of $g_{a\gamma\gamma} \gtrsim 1.3 \times 10^{-9}~\si{\GeV^{-1}}$ for a measurement time of 30 days. Future upgrades, including pressure tuning of the effective refractive index and implementation of a Fabry--Pérot cavity, could extend the accessible mass range and improve sensitivity, establishing WISPFI as a scalable platform to explore previously inaccessible regions of the axion parameter space.

\end{abstract}



\maketitle

\section{Introduction}
\label{section:Introduction}

Axions are hypothetical pseudoscalar particles originally postulated as a consequence of the Peccei–Quinn solution to the strong CP problem in QCD \cite{peccei_mathrmcp_1977,kim_weak-interaction_1979,dine_simple_1981,Sikivie_experimental_1983}. Beyond their role in particle theory, axions and axion-like particles (ALPs) are compelling dark matter (DM) candidates and a target of broad experimental effort \cite{ringwald_review_axions_2024}. The QCD axion mass $m_a$ is related to the axion–photon coupling $g_{a\gamma\gamma} \propto m_a$, enabling photons to convert into axions (and vice versa) in the presence of strong electromagnetic fields \cite{raffelt_mixing_1988,Raffelt_2024}. 

The strongest laboratory limits to date are achieved by haloscope experiments developed over the past 40 years in the \si{\micro\eV} range, which exploit resonant microwave cavities immersed in strong magnetic fields \cite{admx_collaboration_extended_2020,Youn_haloscopes_2024}. 
However, such searches implicitly rely on assumptions on the local axion DM density, which is poorly constrained and may deviate from the canonical galactic average \cite{eggemeier_minivoids_2022}.
 Approaches that do not assume axions as CDM, such as light-shining-through-wall (LSW) experiments or precision polarization measurements, have complementary sensitivity but remain well above the QCD axion band \cite{Cameron_optic_1993,zyla_particle_2020}.
 Moreover, the axion mass range at the level of $\mathcal{O}(100~\mathrm{meV})$ is largely unexplored by direct detection experiments (with the exception of CAST \cite{CAST_Micromegas_2024}), since for cavity-based searches the resonance frequency scales inversely with the cavity volume, making it practically impossible to reach the required frequencies without severely reducing sensitivity.

Here we present the first prototype of WISP Searches on a Fiber Interferometer (WISPFI), a novel laboratory scheme designed to probe photon–axion conversion inside a waveguide. 
The concept is based on a Mach–Zehnder interferometer in which one arm is exposed to a strong transverse magnetic field while the other serves as a reference. 
Operating the interferometer close to a dark fringe minimizes the contribution of laser shot noise (SN) \cite{Gerberding_interferometers_2022}, thereby enhancing sensitivity.
 A photon amplitude reduction in the magnetized arm would indicate photon–axion conversion via the Primakoff effect \cite{Primakoff_1935}. 

In the prototype currently under development at the University of Hamburg, Germany, the sensing arm incorporates a hollow-core photonic crystal fiber (HC-PCF). 
These fibers utilize a photonic bandgap cladding to confine light within a hollow, gas-filled core, where the guided mode can achieve an effective refractive index $n_\mathrm{eff} < 1$.
 This property enables resonant photon–axion conversion at real axion masses, a key requirement that cannot be met with conventional dielectric fibers.
 In the prototype, a $\sim\SI{1}{\meter}$-long HC-PCF segment inside the magnetic field serves as the interaction region, operating at standard atmospheric pressure  conditions and thus probing a fixed axion mass of approximately \SI{50}{\milli\eV}.
 In the envisioned full-scale WISPFI, the probed mass will become tunable in the $\sim$\SIrange{28}{100}{\milli\eV} range by adjusting the gas pressure inside the HC-PCF, which modifies $n_\mathrm{eff}$ of the propagating mode.

This waveguide-based, resonant detection scheme opens a new model-independent experimental window at the \SI{50}{\milli\eV} mass scale and serves as a test-bench for the final design and realization of a future full-scale WISPFI experiment.

\section{Photon-axion mixing and resonant conversion}
\label{section:photon_axion_mixing}

The photon-to-axion conversion probability along the propagation direction $z$ is \cite{Raffelt_2024}:
\begin{equation}
P_{\gamma \rightarrow a} = \sin^{2}(2\theta)\,\sin^{2}(k_\mathrm{osc} z),
\end{equation}
where the mixing angle $\theta$ and oscillation wave number $k_\mathrm{osc}$ determine the amplitude and periodicity of the oscillations, respectively. 
The mixing angle is obtained by diagonalizing the photon-axion mixing matrix with the off-diagonal coupling  term $G = g_{a\gamma\gamma}B/2$ (also known as the mixing energy):  
\begin{equation}
\label{eq:mixing_angle}
\tan(2\theta) = \frac{G}{\Delta_{-}}, \qquad \Delta_{-} = \frac{k_{\gamma}^{2}- k_{a}^{2}}{4\omega_{\gamma}},
\end{equation}
where  $\Delta_{-}$ is the photon-axion transfer momentum, $\omega_{\gamma}$ the photon energy,  $k_{\gamma}^{2}=\omega_{\gamma}^{2}-m_{\gamma}^{2}$ and $k_{a}^{2}=\omega_{a}^{2}-m_{a}^{2}$ are the photon and axion momenta in the medium, respectively.

In the same way, we can define the oscillation wavenumber in terms of these two parameters:
\begin{equation}
\label{eq:kosc}
k_\mathrm{osc}=\sqrt{\Delta_{-}^2+G^2}.
\end{equation}
Under resonant conditions, when the photon and axion momenta match $(k_\gamma = k_a$), the mixing angle reaches $\theta = 45^\circ$, maximizing the conversion probability. 
In this limit, the oscillation wave number reduces to $k_\mathrm{osc} = G$, and for small probabilities ($P_{\gamma \rightarrow a} \ll 1$) the conversion probability simplifies to  
\begin{equation}
P_{\gamma \rightarrow a} \approx \left(\tfrac{1}{2}\, g_{a\gamma\gamma}\, B\, z\right)^{2}.
\end{equation}
The resulting axion mass that can be probed for a given effective refractive index $n_\mathrm{eff}$ of the medium is given by:
\begin{equation}
\label{eq:axion_mass}
    m_a = \omega \sqrt{1 - n_\mathrm{eff}^2}.
\end{equation}

This resonance condition cannot be achieved in waveguides like standard dielectric fibers, which rely on index-guiding mechanisms and have $n_\mathrm{eff} > 1$. 
However, HC-PCFs can guide light through a low-index gas-filled hollow core surrounded by a photonic bandgap cladding with a higher index, resulting in an effective mode index $n_\mathrm{eff} < 1$ \cite{cregan_single_1999, russel_photonic_2003}. 
This property allows real axion masses that satisfy the resonance condition. Combined with their high laser damage threshold \cite{Laurent_2004}, HC-PCFs are an attractive platform for high-sensitivity axion searches \cite{nikodem_laser_2020}.

Based on Eq.~\ref{eq:axion_mass}, the effective mode index of the guided mode, $n_\mathrm{eff}$, determines the axion mass that can be resonantly probed. 
The real part of the propagation constant $k_{\gamma}$ can be analytically approximated via Marcatili’s formula, which is based on a simplified model of a cylindrical hollow‑core fiber \cite{marcatili_hollow_1964}:
\begin{equation}
\label{eq:neff_formula}
n_\mathrm{eff} = \frac{k_\gamma}{k_0} = \sqrt{n_\mathrm{gas}^2(\lambda,p,T) - \left(\frac{u_\mathrm{nm}}{k_0 R_c}\right)^2},
\end{equation}
where $R_c$ is the fiber core radius, $n_\mathrm{gas}$ is the refractive index of the gas filling the core, $\lambda$ is the wavelength, $T$ the temperature, and $p$ the pressure. 
For higher accuracy, especially in fibers with complex geometries, $n_\mathrm{eff}$ can be computed using finite-element method (FEM) simulations that account for the detailed fiber structure.

\begin{figure}[!htb]
    \centering
    \includegraphics[height=0.3\linewidth]{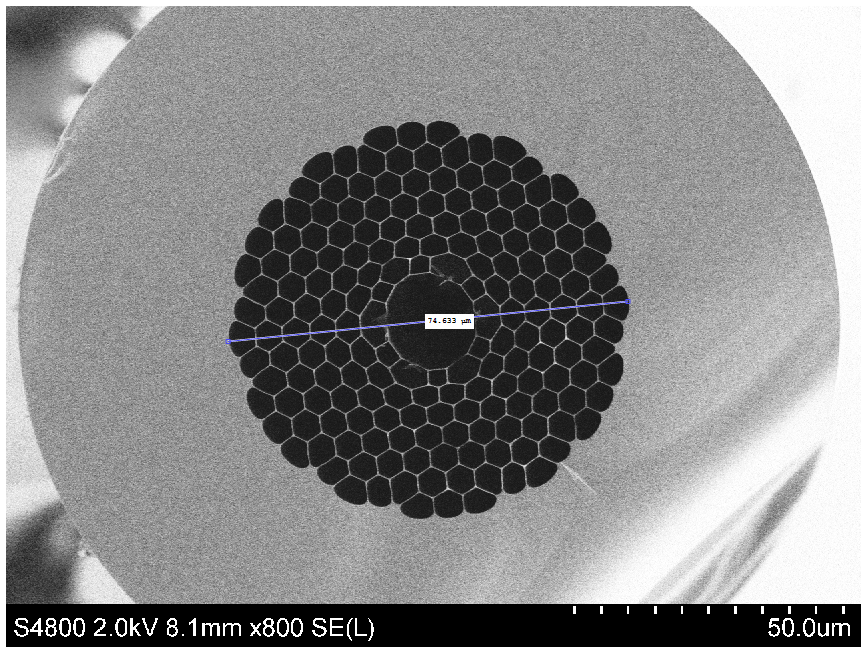}
    \hspace{0.02\linewidth}
    \includegraphics[height=0.3\linewidth]{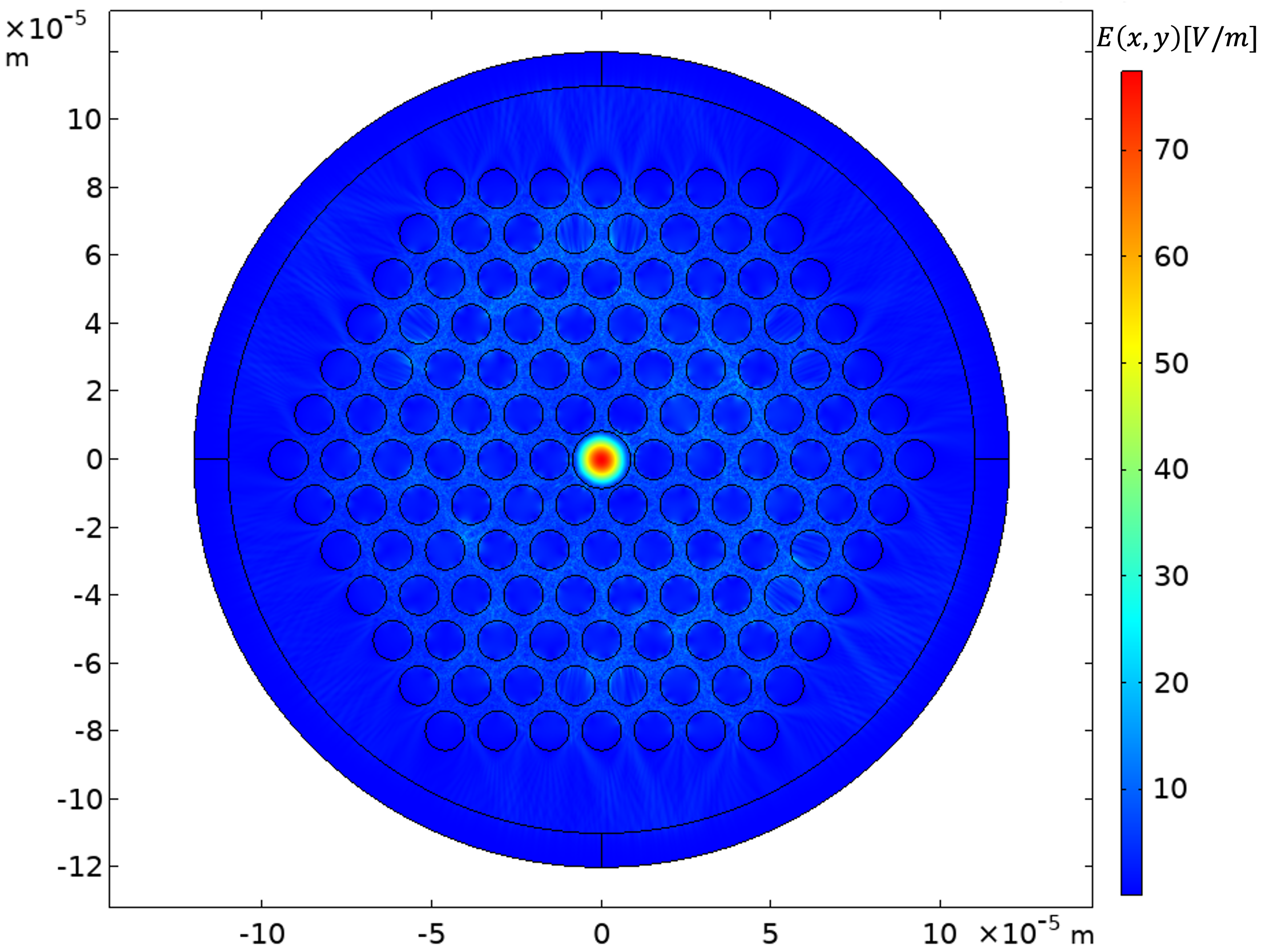}
    \caption{\textbf{Left}: Image taken with a Scanning Electron Microscope (SEM) of a HC-PCF with \SI{8.5}{\micro\meter} core radius. 
    \textbf{Right}: FEM simulation with COMSOL of a mode field distribution in a HC-PCF with \SI{8.5}{\micro\meter} core radius, a capillary-to-core radius ratio of 0.682, and a wavelength of \SI{1.55}{\micro\meter}. The calculated effective mode index is about 0.997 for standard conditions.}
    \label{fig:HC_PCF}
\end{figure}

As we will see in Sect.~\ref{section:Experimental_Setup}, in the WISPFI prototype, the sensing arm incorporates a HC-PCF with a core radius of \SI{8.5}{\micro\meter} (Fig.~\ref{fig:HC_PCF} Left). 
The effective mode index of this fiber is computed using FEM in COMSOL (Fig.~\ref{fig:HC_PCF} Right), taking into account its actual geometry with six rings of air-holes.
 This precise determination of $n_\mathrm{eff}\sim0.997$ sets the specific axion mass that can be probed under standard atmospheric pressure conditions and provides a fixed test point for experimental development.

\section{Principles of phase-modulated interferometry}
\label{section:Principle_of_phase_modulated_interferometry}

The key idea of WISPFI is to detect small axion-induced photon losses by converting them into a measurable modulation at a known frequency \cite{wispfi_2024}. 
Here, we illustrate the principle for the WISPFI prototype, which employs a Mach-Zehnder interferometer (MZI) with a phase-modulated input.\\

Light is injected into Port~1 of the interferometer, while Port~2 remains dark. The input field is:
\begin{equation}
    E_{\text{in1}}(t) = E_0 \, e^{i \omega_\gamma t} \, e^{i \beta_m \sin(\omega_m t)},
\end{equation}
where $\beta_m$ is the modulation index, $\omega_m$ the modulation frequency, and $\omega_\gamma$ the carrier frequency. 
The phase modulation generates sidebands around the carrier, which can be expanded using the Jacobi--Anger identity:
\begin{equation}
    e^{i \beta_m \sin (\omega_m t)} = \sum_{n=-\infty}^{\infty} J_n(\beta_m) e^{i n \omega_m t}.
\end{equation}

The expansion can then be truncated to the carrier and the first-order sidebands ($n=0,\pm1$):
\begin{equation}
    E_{\text{in1}}(t) \approx E_0 e^{i \omega_\gamma t} \Big[J_0(\beta_m) + J_1(\beta_m) e^{i \omega_m t} - J_1(\beta_m) e^{-i \omega_m t}\Big].
\end{equation}

At the first 50:50 beam splitter,
\begin{equation}
    \mathbf{U}_{\rm BS} = \frac{1}{\sqrt{2}} 
    \begin{pmatrix} 1 & i \\ i & 1 \end{pmatrix},
\end{equation}
the field is split into \textit{sensing} ($E_S$) and \textit{reference} ($E_R$) arms:
\begin{equation}
    \begin{pmatrix} E_S \\ E_R \end{pmatrix} 
    = \mathbf{U}_{\rm BS} 
    \begin{pmatrix} E_{\rm in1} \\ 0 \end{pmatrix}
    = \frac{1}{\sqrt{2}} 
    \begin{pmatrix} E_{\rm in1} \\ i E_{\rm in1} \end{pmatrix}.
\end{equation}

The sensing arm has a length $L_S = L + \Delta L/2$ and is exposed to the magnetic field, where photon--axion conversion occurs. The reference arm has a length $L_R = L - \Delta L/2$ and remains unperturbed. After propagation, the fields in the two arms are given by:
\begin{align}
    E_S(t) &= \frac{1}{\sqrt{2}} E_0 (1 - P_{\gamma \to a}) e^{i (\omega_\gamma t - k_\gamma L_S)} 
    e^{i \beta_m \sin(\omega_m t - k_m L_S)}, \\
    E_R(t) &= \frac{i}{\sqrt{2}} E_0 e^{i (\omega_\gamma t - k_\gamma L_R)} 
    e^{i \beta_m \sin(\omega_m t - k_m L_R)},
\end{align}
where $P_{\gamma \to a}$ is the photon--axion conversion probability, $k_\gamma$ is the carrier wavenumber, and $k_m$ the modulation wavenumber.

The two arms are recombined at the second beam splitter:
\begin{equation}
    \begin{pmatrix} E_{\rm out1} \\ E_{\rm out2} \end{pmatrix} 
    = \mathbf{U}_{\rm BS} 
    \begin{pmatrix} E_S \\ E_R \end{pmatrix}
    = \frac{1}{\sqrt{2}} 
    \begin{pmatrix} E_S + i E_R \\ i E_S + E_R \end{pmatrix}.
\end{equation}

To suppress the large carrier background and maximize sensitivity to small amplitude changes, the interferometer is locked close to a dark fringe for the carrier.
This ensures destructive interference of the carrier at the output port, thereby reducing laser SN. 
The dark-fringe conditions are
\begin{align}
    k_{\gamma}\Delta L &= (2q+1)\pi + \alpha_{\gamma}, \quad q \in \mathbb{Z}, \\
    k_m \Delta L &= (2p+1)\pi + \alpha_m, \quad p \in \mathbb{Z},
\end{align}
where $\alpha_{\gamma} = k_{\gamma}\sigma_{\gamma}$ and $\alpha_m = (k_m/k_\gamma)\alpha_{\gamma}$ are small phase errors due to path-length mismatch. In practice, we consider only the carrier phase error $\alpha_{\gamma}$ to be significant.

With these conditions, the power at the dark port (Port~1) becomes
\begin{equation}
\begin{aligned}
    P_{\rm out1}(t) \approx & \; 4 E_0^2 J_1^2(\beta_m) \sin^2(\omega_m t - k_m L + \alpha_{\gamma}) \\
    & - 2 E_0^2 J_0(\beta_m) J_1(\beta_m) P_{\gamma \to a} 
    \sin(\omega_m t - k_m L + \alpha_{\gamma}).
\end{aligned}
\end{equation}
The first term corresponds to a large oscillation from sideband self-interference, while the second is a small modulation proportional to $P_{\gamma \to a}$, arising from carrier--sideband mixing.

Finally, the axion signal is isolated by demodulating $P_{\rm out1}(t)$ at frequency $\omega_m$ (e.g.\ with a lock-in amplifier, see Fig.~\ref{fig:wispfi_prototype_setup}). The time-averaged demodulated signal is:
\begin{equation}
    P_{\text{av}} = 2E_0^2 J_0(\beta_m) J_1(\beta_m) P_{\gamma \rightarrow a},
\end{equation}
which separates the axion-induced contribution from the background noise.
In practice, the axion signal detection involves comparing the spectra recorded by the dark-port photodiode over long integration times between the $B = \mathrm{ON}$, where photon-axion conversion occurs, and $B = \mathrm{OFF}$, where no conversion is present. 
This demonstrates how phase modulation and dark-fringe operation together convert a small photon--axion conversion probability into a measurable signal.

\section{Experimental setup}
\label{section:Experimental_Setup}

\begin{figure}[!htb]
    \centering
    \includegraphics[width=\linewidth]{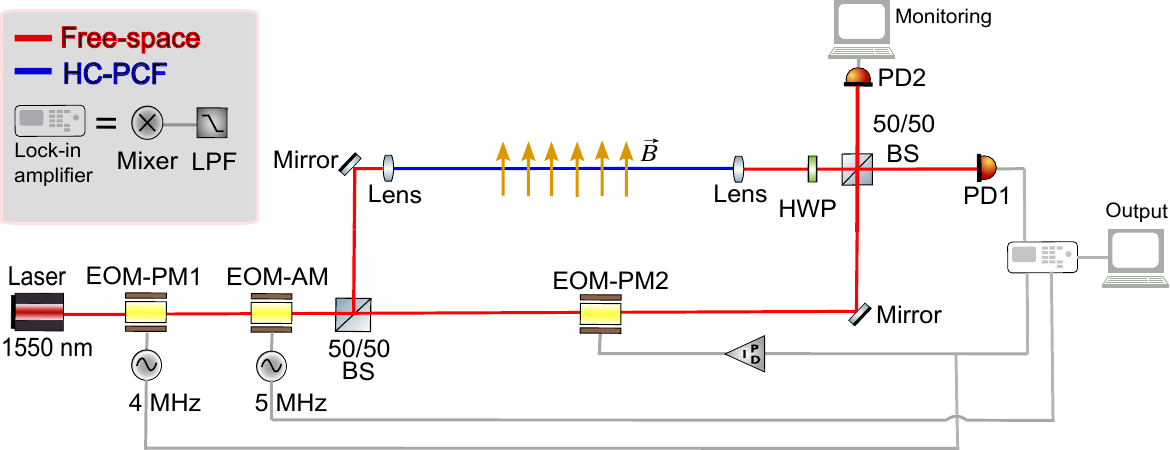}
    \caption{
Schematic of the WISPFI prototype using a partial free-space MZI for axion-photon oscillation detection. 
The free-space laser beam is shown in red, while the HC-PCF arm inside the magnetic field (blue) represents the sensitive path for photon-axion conversion. 
The interferometer uses a resonant EOM-PM at \SI{4}{\mega\hertz} to provide the phase-lock reference, a broadband EOM-PM to compensate phase drifts, and a resonant EOM-AM at \SI{5}{\mega\hertz} for axion-induced amplitude detection. 
Other components include beam splitters (BS), a half-wave plate (HWP), photodiodes (PD), and a low-pass filter (LPF).}
    \label{fig:wispfi_prototype_setup}
\end{figure}

The WISPFI prototype is designed to detect small axion-induced photon amplitude changes using a partial free-space MZI.
As discussed in Sect.~\ref{section:Principle_of_phase_modulated_interferometry}, phase modulation allows the interferometer to be locked near the dark fringe, while a dedicated amplitude modulation frequency can be used to isolate axion-induced amplitude changes from other noise sources. 
A schematic of the experimental setup is shown in Fig.~\ref{fig:wispfi_prototype_setup}.

The interferometer is driven by a high-power \SI{2}{\watt}, \SI{1550}{\nano\meter} laser. 
The beam initially passes through a resonant phase electro-optical modulator (EOM-PM1) operating at \SI{4}{\mega\hertz}, which provides the frequency reference for phase-locking the interferometer close to the dark fringe. 
A separate broadband phase EOM (EOM-PM2) in the sensing arm compensates for phase variations between the two interferometer arms.
 The error signal from the dark-port photodiode (PD1) is demodulated at the phase-lock frequency of \SI{4}{\mega\hertz} and low-pass filtered with a \SI{10}{\hertz} bandwidth before being fed into a PID loop controlling the broadband EOM.
This stabilizes both the phase and the amplitude working point near the dark fringe.

After the EOM-PM, a resonant amplitude modulator (EOM-AM) operating at \SI{5}{\mega\hertz} is used to provide a dedicated frequency for measuring axion-induced amplitude changes. 
The signal is demodulated at this frequency using a local oscillator and the same low-pass filter as used for the phase lock.
This frequency choice ensures that the phase-lock system does not compensate for amplitude changes caused by photon-to-axion conversion.

The laser beam is then split at a 50:50 beam splitter (BS) into a reference and a sensing arm.
The sensing arm contains a \SI{1}{\meter}-long HC-PCF with a core radius of \SI{8.5}{\micro\meter}, provided by MPI Erlangen \cite{Uebel_2016} (see Fig.~\ref{fig:HC_PCF}).
Free-space coupling into the HC-PCF is achieved with a stable efficiency of about $80\%$, marking a first milestone that can be further improved with optimized mode matching.

The fiber will be mounted in a custom-made aluminum holder with a production accuracy of approximately \SI{20}{\micro\meter}, ensuring that it remains straight so that the full conversion length contributes to the measurement. 
A half-wave plate (HWP) positioned after the HC‑PCF fine‑tunes the output polarization, ensuring optimal alignment and thereby minimizing the modulation effort required from the broadband EOM.

\begin{figure}[!htb]
    \centering
    \includegraphics[width=0.57\linewidth]{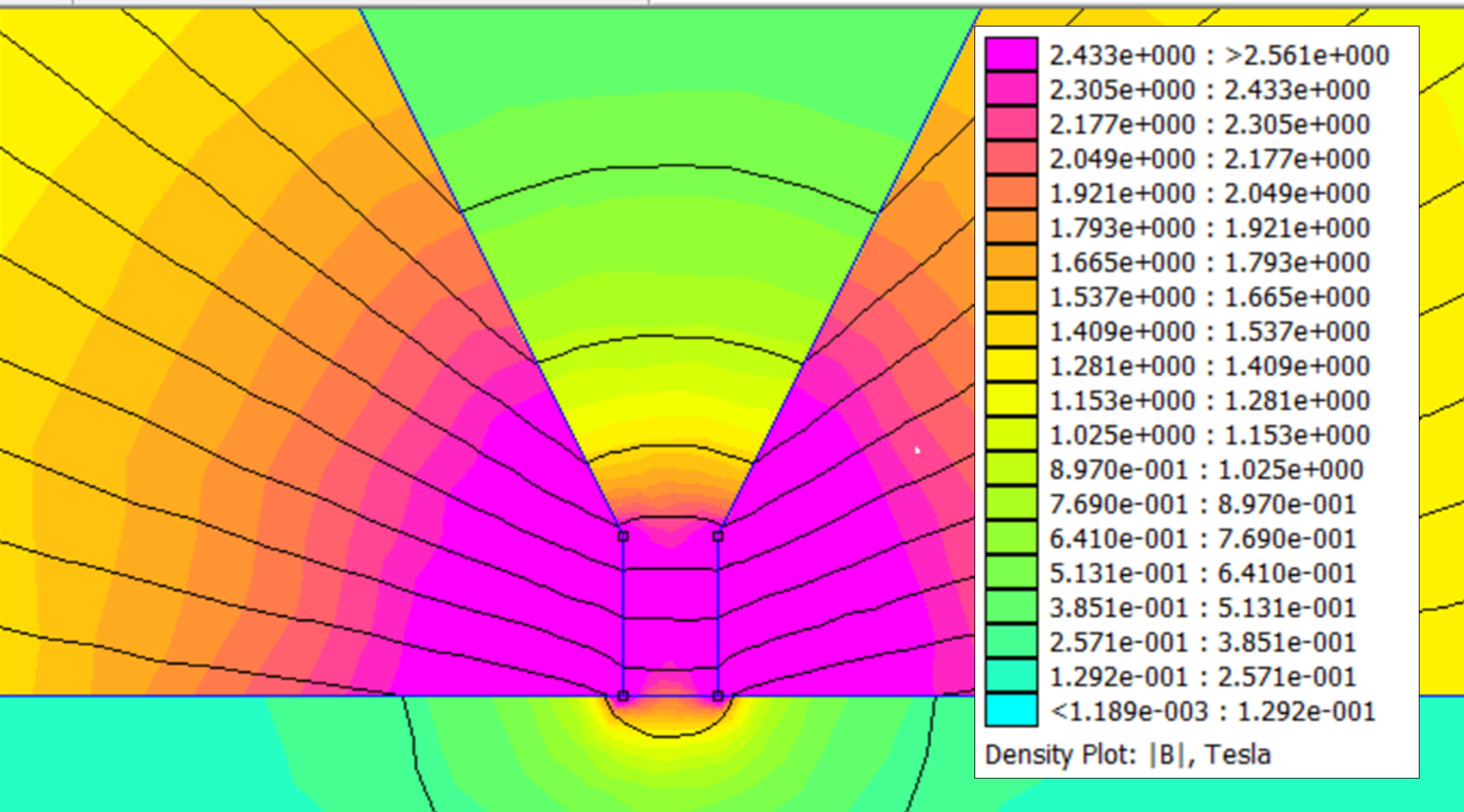}
    \includegraphics[width=0.35\linewidth]{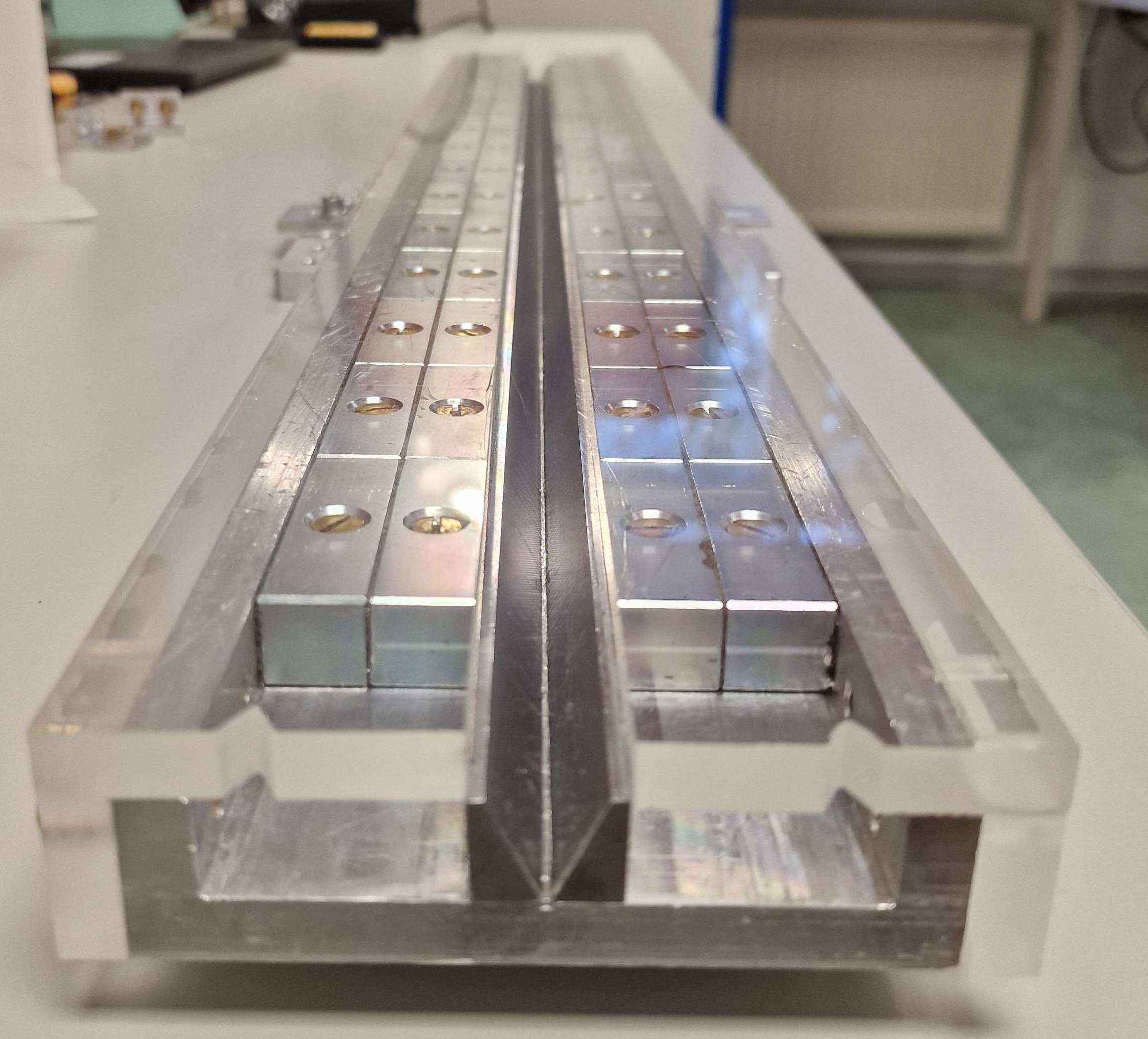}
        \caption{
\textbf{Left}: Finite-element simulation of the magnetic field in the WISPFI prototype.
The field intensity is shown in the center of the gap between two Nd permanent-magnet arrays separated by \SI{0.6}{\milli\meter}, through which the HC-PCF is threaded. 
The maximum field in this middle region reaches approximately \SI{2}{\tesla}. 
Replacing the Fe wedge with a Co-Fe alloy would increase the field up to \SI{3.7}{\tesla}. 
\textbf{Right}: Picture of the actual magnet panel used in the WISPFI prototype setup.}
    \label{fig:wispfi_magnet_simulation}
\end{figure}

Photon-axion conversion occurs in an external magnetic field generated by a custom array of 60 Nd permanent magnets. 
To increase the magnetic flux density in the region of the fiber, a removable thin Fe wedge is positioned between the two magnet arrays, creating a \SI{0.6}{\milli\meter} gap through which the HC-PCF is threaded (see Fig.~\ref{fig:wispfi_magnet_simulation} right). 
Finite-element simulations of the magnetic field in this region are shown in Fig.~\ref{fig:wispfi_magnet_simulation} left, and indicate a field of approximately \SI{2}{\tesla} at the location of the HC-PCF. 
Replacing the Fe wedge with a Co-Fe alloy could further increase the field up to \SI{3.7}{\tesla}. 
The magnet assembly is mounted on a motorized stage that vertically moves the magnet panel, positioning it under the HC-PCF for the $B=\SI{2}{\tesla}$ state or retracting it away for the $B=\SI{0}{\tesla}$ state. 
This allows direct comparison of measurements with and without the magnetic field for identification of axion-induced amplitude modulation.

After recombination at a second 50:50 BS, the dark-port signal in PD1 is used to measure axion-induced amplitude changes, while the bright port (PD2) monitors the overall interferometer performance.

\begin{figure}[!htb]
    \centering
    \includegraphics[width=0.8\linewidth]{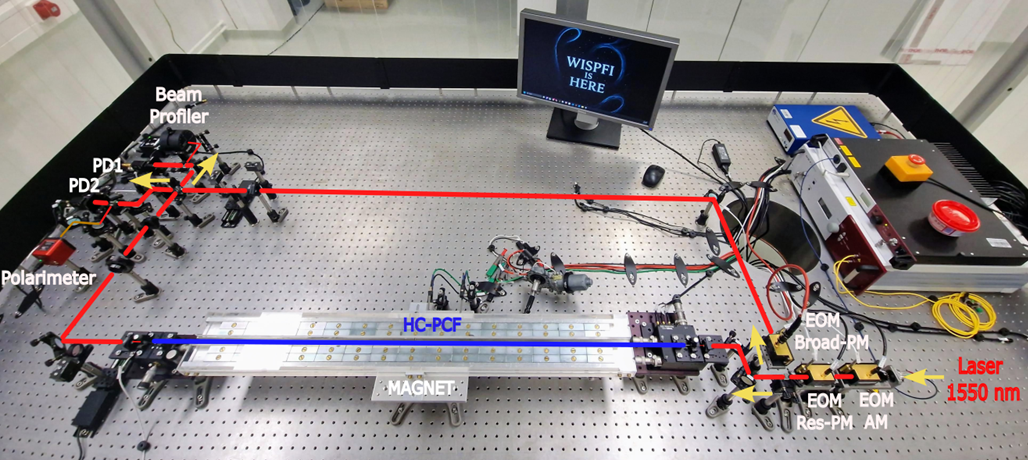}
    \caption{
Photograph of the WISPFI prototype under commissioning at the University of Hamburg. 
Free-space optical paths are indicated in red, while the HC-PCF in the sensing arm is shown in blue inside the magnet array. 
The aluminum fiber holder ensuring straight alignment is not yet implemented in this picture.}
    \label{fig:wispfi_prototype_setup_photo}
\end{figure}

Finally, a fully automated Python-based data acquisition (DAQ) system has been designed that monitors multiple system parameters, including polarization, beam profile, laser power, laser wavelength, temperature, and humidity. 
These measurements will provide the quality criteria for the analysis, ensuring stable conditions and reproducible measurements throughout the experiment. 
A photograph of the current laboratory implementation of the WISPFI prototype is shown in Fig.~\ref{fig:wispfi_prototype_setup_photo}.

\section{Projected Sensitivity}
\label{section:Sensitivity}

The sensitivity of the WISPFI prototype is evaluated assuming operation near a dark fringe, where the interferometer converts small axion-induced photon losses into measurable amplitude modulations. 
In this regime, the main noise sources are the dark current of the photodiode at the dark port (PD1) and the laser SN.

As described in Sect.~\ref{section:Experimental_Setup}, the interferometer uses a high-power \SI{2}{\watt}, \SI{1550}{\nano\meter} laser coupled to a \SI{1}{\meter}-long HC-PCF placed in a \SI{2}{\tesla} magnetic field. 
Phase-locking of the interferometer is achieved near the dark fringe with a residual power of 1\% at PD1. Under these conditions, the noise-equivalent power (NEP) contributions are $\text{NEP}_{\rm SN} = \SI{0.11}{\femto\watt/\sqrt{\hertz}}$ for SN and $\text{NEP}_{\rm PD1} \approx \SI{2}{\femto\watt/\sqrt{\hertz}}$ for the InGaAs photodiode dark current. 
Hence, the dominant noise source in the measurement is the PD1 dark current.

The expected sensitivity of the prototype allows probing an axion-photon coupling of
$g_{a\gamma\gamma} \gtrsim 1.32 \times 10^{-9}~\si{\GeV^{-1}}$ for an axion mass of $m_a = 4.93 \times 10^{-2}~\si{\eV}$ over a 30-day measurement period. 
This sensitivity can be expressed as:
\begin{equation}
\label{eq:sensitivity_baseline_setup}
\begin{aligned}
    g_{a\gamma\gamma} &\approx 1.32\times10^{-9}\si{\GeV\tothe{-1}} 
    \left(\frac{\mathrm{SNR}}{3}\right)^{1/2}
    \left(\frac{B}{\SI{2}{\tesla}}\right)^{-1} 
    \left(\frac{L}{\SI{1}{\meter}}\right)^{-1} \\
    &\quad \times \left(\frac{P_\mathrm{tot}}{\SI{2}{\watt}}\right)^{-1/2}
    \left(\frac{\beta_\mathrm{phase}}{1}\right)^{-1/2}
    \left(\frac{t}{\SI{30}{\day}}\right)^{-1/4} \\
    &\quad \times \left(\frac{\mathrm{NEP_{SN+PD1}}}{\SI{2.11}{\femto\watt\per\sqrt{\Hz}}}\right)^{1/4}
    \left(\frac{\Delta{\nu}}{\SI{10}{\Hz}}\right)^{1/8}
    \left(\frac{\mathrm{E_{ph}}}{\SI{0.8}{\eV}}\right)^{1/4}.
\end{aligned}
\end{equation}
Here, the phase modulation factor is defined as $\beta_{\rm phase} = 2 J_0(\beta_{\rm m}) J_1(\beta_{\rm m})$. 
For a modulation index $\beta_{\rm m} = 1$, $J_0(1) \approx 0.7652$ and $J_1(1) \approx 0.4401$, giving $\beta_{\rm phase} \approx 0.67$.

The HC-PCF is assumed to remain straight, with its electric field aligned to the magnetic field to maximize photon-axion conversion. 
The effective mode index of the propagating mode is computed assuming a constant temperature of \SI{20}{\celsius} and a pressure of \SI{1}{\bar} along the \SI{1}{\meter}-long fiber. 
As discussed in Sect.~\ref{section:photon_axion_mixing}, the resonant axion mass (Eq.~\ref{eq:axion_mass}) is primarily determined by the fiber core radius $R_c$ (Eq.~\ref{eq:neff_formula}).

The HC-PCF manufacturing process introduces small random variations in the core radius along the fiber, which can be characterized as low-frequency ($1/f$-like) fluctuations with typical length scales longer than the fiber segment. 
Larger variations in the core radius would broaden the accessible axion mass range while reducing the peak sensitivity. 
To account for these variations, we consider an average core radius $R_c = \SI{8.5}{\micro\meter}$ with a standard deviation $\sigma = \SI{10}{\nano\meter}$. 
For the sensitivity estimate, ten \SI{1}{\meter}-long fiber realizations are simulated, each divided into 100 segments of \SI{0.01}{\meter}, and the photon-to-axion conversion probability $P_{\gamma \rightarrow a}$ is computed using a transfer-matrix approach \cite{mirizzi_stochastic_2009, deangelis_relevance_2011}.  

\begin{figure}[!htb]
\centering
\includegraphics[width=0.85\linewidth]{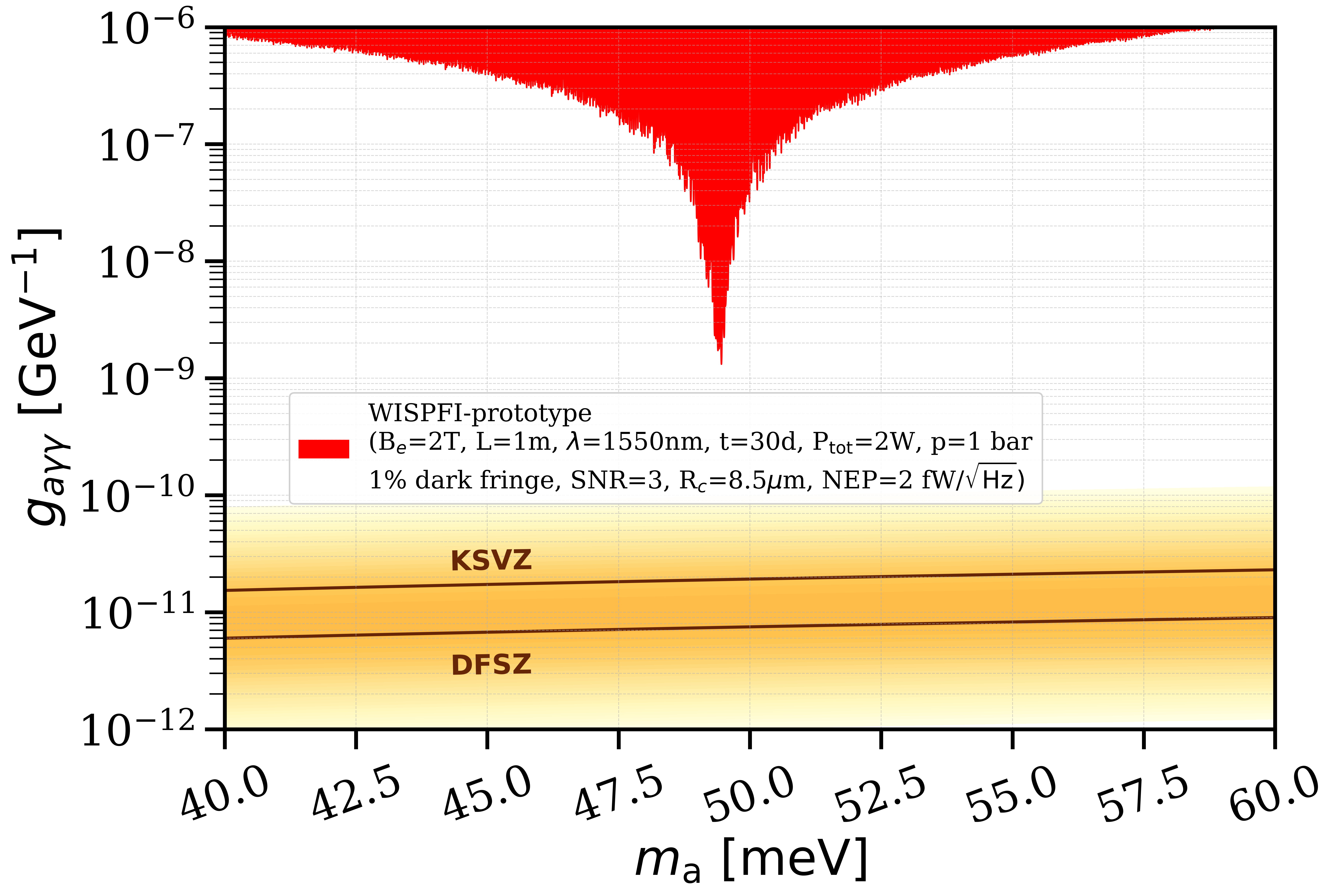}
\caption{Projected median sensitivity of the WISPFI prototype assuming a \SI{2}{\tesla} magnetic field. 
A \SI{2}{\watt} laser is coupled into a \SI{1}{\meter}-long HC-PCF with an average core radius of \SI{8.5}{\micro\meter} and $\sigma = \SI{10}{\nano\meter}$. 
The laser wavelength is \SI{1.55}{\micro\meter}, and a total measurement time of 30~days is assumed.}
\label{fig:wispfi_prototype_exclusion_plot}
\end{figure}

Fig.~\ref{fig:wispfi_prototype_exclusion_plot} shows the resulting median sensitivity of the WISPFI prototype.

\section{Summary and Discussion}

The WISPFI prototype represents the first table-top, DM-independent experiment capable of probing ALPs in a previously unexplored mass range around \SI{50}{\milli\eV}. 
Using a HC-PCF in a phase-locked MZI, combined with a permanent Nd magnet array, the setup demonstrates the feasibility of detecting photon-ALP conversion in a compact interferometric platform.
 With the current configuration of \SI{1}{\meter} of HC-PCF, \SI{2}{\tesla} magnetic field, and \SI{2}{\watt} laser power, the experiment is expected to probe an axion-photon coupling of $g_{a\gamma\gamma} \gtrsim 1.3 \times 10^{-9}~\si{\GeV^{-1}}$ over a measurement period of 30 days, providing a unique platform to explore this otherwise inaccessible parameter space. 
Phase-locking the interferometer near the dark fringe and using a dedicated amplitude modulation frequency ensures high-precision and reliable ALP signal detection.

Beyond the current prototype, several strategies have been identified to improve sensitivity and expand the mass coverage. 
Implementing a controlled pressure-tuning procedure in the HC-PCF \cite{cao_fiber_2014, Masum_fiber_2019}, would allow systematic scanning of the resonant axion mass by modifying the effective refractive index of the guided mode. 
Additionally, integrating a Fabry-Pérot cavity along the sensing arm could further enhance the effective conversion length and optical power.
Such cavities with finesses exceeding \num{3000} have been demonstrated using Pound-Drever-Hall locking, providing a practical route to boost the signal \cite{Ding_fabry_2020, tam_production_2012}.

Overall, the prototype validates the core experimental principles of the WISPFI concept and provides a versatile test-bench for the full-scale experiment. 
The combination of strong magnetic fields, precise phase and amplitude modulation, and automated data acquisition establishes a robust, scalable, and model-independent platform to explore previously inaccessible ALP parameter space.

\section*{Acknowledgments}

JMB, MM acknowledge funding and DH support by the Deutsche Forschungsgemeinschaft (DFG, German Research Foundation) under Germany’s Excellence Strategy – EXC 2121 ``Quantum Universe" – 390833306, and through the DFG funds for major instrumentation grant DFG INST 152/824-1. This article is based upon work from COST Action COSMIC WISPers CA21106, supported by COST (European Cooperation in Science and Technology).



\bibliography{references}

\end{document}